\documentclass[prl,amsmath,amssymb,twocolumn,unsortedaddress]{revtex4-1}
\usepackage{style}
\usepackage{graphicx}  
\usepackage{dcolumn}   
\usepackage{bm}        
\usepackage{float}
\usepackage{multirow}
\usepackage{amsmath}
\usepackage{amsmath,amssymb}
\usepackage{epsfig}
\usepackage{feynmp}
\usepackage{hyperref}
\usepackage{color}
\hypersetup{
	linktoc=all
	}

\def\bk{\mathbf{k}}

\def\cH{{\mathcal H}}

\def\cZ{{\mathcal Z}}

\def\mR{{\mathbb{R}}}

\def\mZ{{\mathbb{Z}}}

\def\ul{\underline}

\def\tw{{\tilde{\omega}}}

\def\btimes{\hbox{\boldmath $\times$}}

\def\be{\begin{equation}}
\def\ee{\end{equation}}
\def\lb{\label}

\begin{document}
\setcounter{totalnumber}{10}

\title{Wave-Turbulence Theory of four-wave nonlinear interactions}

\author{Sergio~Chibbaro\textsuperscript{a},  Giovanni Dematteis\textsuperscript{b}, Christophe Josserand\textsuperscript{a,c} and Lamberto Rondoni\textsuperscript{b,d,e}}
\affiliation{\textsuperscript{a}Sorbonne Universit\'e, UPMC Univ Paris 06, CNRS, UMR 7190, Institut Jean Le Rond d'Alembert, F-75005 Paris, France\\
\textsuperscript{b}Dipartimento di Scienze Matematiche, Politecnico di Torino, Corso Duca degli Abruzzi 24, I-10129 Torino, Italy;\\
\textsuperscript{c}Laboratoire d'Hydrodynamique (LadHyX), UMR 7646 CNRS-Ecole Polytechnique, 91128 Palaiseau CEDEX, France
\textsuperscript{d}INFN, Sezione di Torino, Via P. Giuria 1, I-10125, Torino, Italy\\
\textsuperscript{e}MICEMS, Universiti Putra Malaysia, 43400 Serdang Selangor, Malaysia}

\begin{abstract}
The Sagdeev-Zaslavski (SZ) equation for wave turbulence is analytically derived, both in terms of generating 
function and of multi-point pdf, for weakly interacting waves with initial random phases. 
When also initial amplitudes are random, the one-point pdf equation is derived. Such 
analytical calculations remarkably agree with results obtained in totally different fashions.
Numerical investigations of the two-dimensional nonlinear Schr\"odinger equation (NLSE) 
and of a vibrating plate prove that:
(i) generic Hamiltonian 4-wave systems rapidly attain a random distribution of phases independently of the slower dynamics of the amplitudes, vindicating the hypothesis 
of initially random phases;
(ii) relaxation of the Fourier amplitudes to the predicted stationary distribution (exponential) happens on a faster timescale than relaxation of the spectrum 
(Rayleigh-Jeans distribution);
(iii) the pdf equation correctly describes dynamics under different forcings: the NLSE has
an exponential pdf corresponding to a quasi-gaussian solution, like the vibrating plates, that
also show some intermittency at very strong forcings.
\end{abstract}

\maketitle

\textit{Introduction}
Dispersive waves are ubiquitous in nature, and their
nonlinear interactions make them intriguing
and challenging \cite{whitham2011linear,berry2000making}.
Wave Turbulence is the theory that describes the statistical properties of large numbers of 
incoherent interacting waves, with tools such as the {\it wave kinetic equation} analytically 
derived in the late sixties.
This equation describes the evolution of the wave spectrum in time, when homogeneity and weak nonlinearity are assumed
\cite{falkovich1992kolmogorov,nazarenko2011wave,newell2011wave}. It has been applied
to numerous phenomena, including ocean waves \cite{komen94,onorato2002freely,falcon2007observation}, capillary waves 
\cite{pushkarev1996turbulence,falcon2009capillary} Alfv\'en waves \cite{galtier2000weak}, optical waves \cite{picozzi2014optical} and solid oscillations \cite{Dur_06,Mor_08,Bou_08,Miq_13,Cad_13,Humbert16}.
It is the analogue of the Boltzmann equation for 
classical particles and it allows the Rayleigh-Jeans equilibrium state as well as non-equilibrium solutions, in terms 
of Kolmogorov-Zakharov (KZ) spectra \cite{zakharov1967energy}. 

To characterise the invariant measure of the dynamics, that is to find the complete statistical 
description concerning all quantities of interest, an important step has been taken by Sagdeev 
and Zaslavski \cite{zaslavskii1967limits}, who obtained the Brout-Prigogine
equation for the probability density function (pdf) of wave turbulence \cite{brout1956statistical}.
More recently, this statistical framework has been nicely revisited 
using the diagrammatic technique~\cite{nazarenko2011wave} and performing analytical calculations, in the $3$-wave case~\cite{Cho_05a,choi2005joint,Eyi_12}.
Interestingly, many experimental and theoretical results have shown that deviations from wave-turbulence 
predictions can be found for rare events, {\em e.g.}\ \emph{intermittency} \cite{majda1997one,falcon2007observation,falcon2008fluctuations,lukaschuk2009gravity,nazarenko2010statistics,falcon2010origin}. This seems to be  the case when a more general theoretical framework \cite{New_01a,New_01b,connaughton2003dimensional,Lvo_04,jakobsen2004invariant} is required, because the nonlinearities are not small \cite{chibbaro2015elastic,chibbaro2016weak}.

In this  communication, the complete wave-turbulence theory is developed for a fully general 4-wave system, 
whose hamiltonian is expressed by the following canonical expression:
\begin{equation}\label{eq: H}
H=\frac{1}{2}\sum_{1}\omega_{1} A^{\sigma_1}_{1} A^{-\sigma_1}_{1} + \epsilon \sum_{1234} \mathcal{H}^{\underline{\sigma}}_{\underline{\mathbf{k}}}\:A_1^{\sigma_1}A_2^{\sigma_2}A_3^{\sigma_3} A_4^{\sigma_4} \:\delta_{\underline{\sigma}\cdot \underline{\mathbf{k}},\mathbf{0}}~.
\end{equation}
Here, $\omega_{1}$ is the normal frequency of wave $1$, that nonlinearly interacts with waves $2,3,4$ with 
coupling constant $\mathcal{H}^{\underline{\sigma}}_{\underline{\mathbf{k}}}$, $\sum_i\doteq\sum_{\sigma_i=\pm1}\sum_{\bk_i\in\Lambda_L^*}$, $\Lambda^*_L=\frac{2\pi}{L}\mathbb{Z}^d_M$. $A_\bk^{\sigma}=\frac{1}{\sqrt{2}}(P_\bk + i\sigma Q_\bk)$ are the canonical variables of the wave-field, whose real and imaginary parts are the coordinates and momenta. $\sigma=\pm 1$ represents the ``spin" of a wave, so that $A_\bk^+\doteq A_\bk$, $A_\bk^-\doteq A_\bk^*$ (* is complex conjugation).

\textit{Theory}
Given the Hamiltonian (\ref{eq: H}), we concisely derive the equations of motion in terms of canonical normal variables;
the details are given in Ref.\cite{chibbaro20164}. First,
recall that the action-angle variables (amplitudes and phases) for the linear dynamics are defined by 
$J_{\bk}=|A_{\bk}^\sigma|^2$ and $\varphi_\bk=\sigma \arg(A_\bk^\sigma),$
so that $A_{\bk}^\sigma=\sqrt{J_{\bk}}\psi_{\bk}^\sigma$, where $\psi_{\bk}=\exp(i\varphi_{\bk})$. Then,
the Liouville measure $\mu$ preserved by the Hamiltonian flow reads: 
$d\mu=\prod_\bk dQ_\bk dP_\bk = \prod_\bk \frac{1}{i}dA^+_\bk dA^-_\bk 
= \prod_\bk \frac{1}{i}da^+_\bk da^-_\bk =\prod_\bk dJ_\bk d\varphi_\bk. \;$
$A^\sigma_\bk$ and $a^\sigma_\bk$ 
are linked by the rotation in the complex plane: $A_\bk^\sigma = a_\bk^\sigma e^{i\sigma\omega_\bk t}$.
The equations of motion with 4-wave interactions can thus be expressed by ($\sigma=+1$ when it is omitted):
\begin{eqnarray} \lb{eq: after-int-repr}
\frac{\partial a_1}{\partial t}&=&\epsilon \sum_{234}\mathcal{L}^{+ \sigma_2 \sigma_3 \sigma_4}_{1234}a_2^{\sigma_2}a_3^{\sigma_3} a_4^{\sigma_4} \nonumber\\
&& \btimes \exp \left[i \left(-\omega_1+ \sigma_2 \omega_2 + \sigma_3 \omega_3 + \sigma_4 \omega_4 \right) t\right] \\
&& \btimes
\delta_{- \mathbf{k}_1+\sigma_2 \mathbf{k}_2+\sigma_3 \mathbf{k}_3+\sigma_4 \mathbf{k}_4,\mathbf{0}} \nonumber
\end{eqnarray}
For a system with $N$ modes in a box of size $L$, the complete statistical description of the field is given by the \textit{generating function}, defined by:
\be \cZ_L[\lambda,\mu,T]\doteq \left \langle \exp \left( \sum_{\bk \in \Lambda_L^*}\lambda_\bk J_\bk(T) \right) 
\prod_{\bk \in \Lambda_L^*} \psi_\bk^{\mu_\bk}(T)\right \rangle ~,
\label{eq: genfunct}
\ee
where $\lambda_\bk \in \mR,\;\; \mu_\bk \in \mZ$, $\forall \bk \in \Lambda^*_L$.

Assuming that the canonical wavefield enjoys the \textit{Random Phase} (RP) property at the initial time,
we have averaged over phases using the \textit{Feynman-Wyld diagrams} \cite{nazarenko2011wave}.
Further, taking the large-box limit, we have normalized the amplitudes in such a way that the wave spectrum 
remains finite. This step is crucial for the evaluation of the different diagrams~\cite{Eyi_12}. 
Then, taking the large-box limit, followed by the \textit{small nonlinearity} limit,
and introducing the nonlinear time $\tau=\epsilon^2 T$,
we have formally obtained the following closed equation for the generating function (the 
{\em characteristic functional}):
\begin{eqnarray}
&&\frac{d\cZ[\lambda,\mu,\tau]}{d\tau}= -192\pi \delta_{\mu,0}  \nonumber \\
&&\btimes \sum_{\ul{\sigma}}\int d^dk_1d^dk_2d^dk_3d^dk_4 \lambda\left(\bk_1\right)|\cH_{1234}^{-\sigma_2\sigma_3\sigma_4}|^2\delta(\tw^1_{234}) \label{concl1} \\
&& \btimes\delta^1_{234}\bigg(\frac{\delta^3 \cZ}{\delta\lambda(\bk_2)\delta\lambda(\bk_3)\delta\lambda(\bk_4)}-\sigma_2\frac{\delta^3 \cZ}{\delta\lambda(\bk_1)\delta\lambda(\bk_3)\delta\lambda(\bk_4)}+ \nonumber \\
&&\quad-\sigma_3\frac{\delta^3 \cZ}{\delta\lambda(\bk_1)\delta\lambda(\bk_2)\delta\lambda(\bk_4)}-\sigma_4\frac{\delta^3 \cZ}{\delta\lambda(\bk_1)\delta\lambda(\bk_2)\delta\lambda(\bk_3)}\bigg) \nonumber ~,
\end{eqnarray}
which constitutes the main ingredient of the present communication.
The frequency in $\delta(\tw^1_{234})$ has been renormalised~\cite{nazarenko2011wave} as
$\tw_\bk \doteq \omega_\bk+\Omega_\bk$, taking into account the self-interactions possible in 4-wave systems, 
that do not contribute to the nonlinear interactions but shift the linear frequency.

The characteristic functional constitutes the most detailed description of the phenomenon \cite{Monin}, for which
the following holds:
(i) the RP property of the initial field is preserved in time, implying the validity of eq.(\ref{concl1}) for $\tau>0$;
(ii)  eq.(\ref{concl1}) has a solution preserving in time the stricter \textit{Random Phase and Amplitude} (RPA)
property of an initial wavefield, {\em i.e.}\ the possible factorization of $\cZ[\lambda,\mu,0]$;
(iii) differentiating with respect to the $\lambda_\bk$'s, the \textit{spectral hierarchy} for the moments,  
analogous to the BBGKY hierarchy in Kinetic Theory, is obtained. Then, RPA allows us to close the 
hierarchy, leading to the wave spectrum equation, the \textit{kinetic equation}.

\begin{figure}[h]
\includegraphics[scale=0.35]{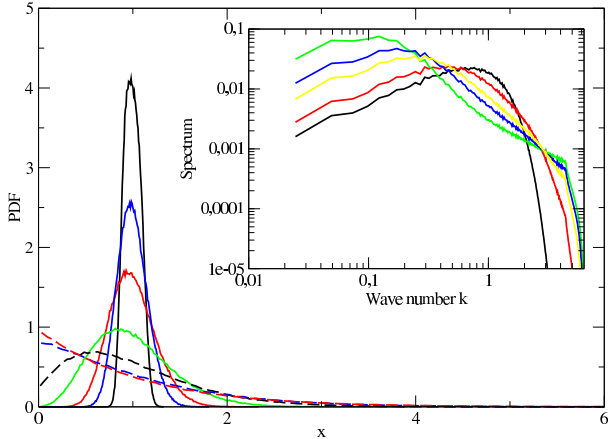}
\caption{Normalized pdf of the modes $|\Psi_{{\bf k}}(\tau)|^2$ for $|{\bf k}|=2$ as a function of the normalized quantity $x=|\Psi_{k}(\tau)|^2/n(k,\tau)$ where $n(k,\tau)$ is the mean value of $|\Psi_{k}(\tau)|^2$. The numerical simulation of the 2D NLSE is performed over a domain of size $256 \times 256$ using a regular square grid of mesh size $dx=0.5$ so that $512\times 512$ modes are simulated. The statistics and mean values are obtained both by an ensemble average over $128$ realizations of the numerical simulation of the NLS equation starting at $\tau=0$ with a Gaussian Fourier mode distribution with random phases, and using the isotropy of the fields allowing angular mean. The pdf are shown for $\tau=0.01$, $0.03$, $0.05$, $0.1$, $0.2$, $0.5$ and $1$ time units respectively from top to bottom. The short time pdf are concentrated around the mean value while they converge at large time to the expected $e^{-x}$ law (corresponding to the dashed red line, pdf for $\tau=10$) and no more variations of the pdf are observed for $\tau>10$. The inset shows the spectrum $n(k,\tau)$ for the times $\tau=0.1$, $10$, $30$, $50$ and $110$, from bottom to top respectively looking at low $k$. 
The equipartition of energy spectrum $n(k,\tau)\propto 1/k^2$ is still not reached for the latest time shown here.}
\label{fig1}
\end{figure}

\begin{figure}[h]
\begin{center}
\includegraphics[scale=0.35]{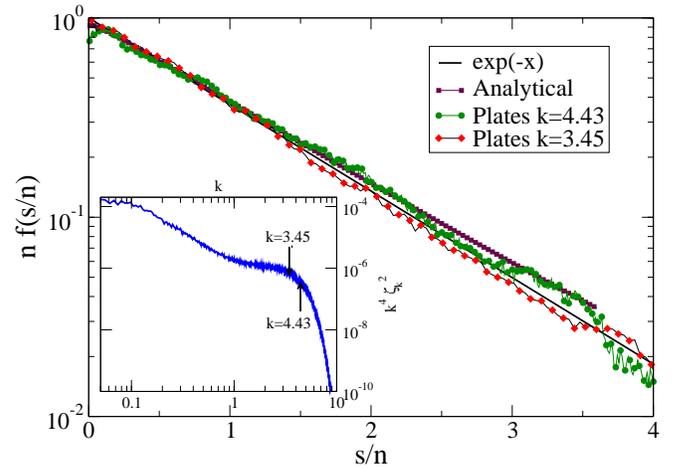}
\caption{Normalized pdf of the Fourier modes $|\zeta_k|^2$ as a function of the rescaled parameter $s/n$ for two different wave numbers $k=3.45$ and $k=4.43$, in a linear-log plot. The statistical average is made using angle average due to the isotropy of the system and time average, because of the statistically stationary regime reached in time. Here $dx=0.25$ and the the square plate is $L\times L=1024\times 1024$, meaning that $4096\times4096$ modes are simulated. The pdf are reasonably well fitted by the equilibrium law $e^{-x}$ although for $k=4.43$ the generalized function (\ref{gensol}) with the cut-off $s_{nl}=3.6\,n(k)$ is a much better fit. The inset shows the compensated spectrum $k^4|\zeta_k|^2$ that exhibits a complex inertial regime, with a $k^{-2}$ slope at large scale ($k \lesssim 1$) indicating intermittent behavior, and the expected weak turbulence spectrum $|\zeta_k|^2 \propto k^{-4}$ at smaller scales ($1 <k <5$), where are located the two modes shown here~\cite{chibbaro2015elastic}. The other modes pdf's show, outside of the forcing region ($k<0.05$), the exponential Rayleigh distribution.}
\label{fig2}
\end{center}
\end{figure}

As the characteristic functional gives too detailed information, in relevant situations
we have derived the equation for the \textit{characteristic function} $\cZ^{(M)}$, 
that concerns a number $M$ of modes, and enjoys the same properties of $\cZ[\lambda,\mu,\tau]$\cite{chibbaro20164}.
Then, under the RPA hypothesis, we derived a closed fully general equation for the {\em1-mode pdf} that reads~\cite{chibbaro20164}:
\be
\frac{\partial P}{\partial \tau}=-\frac{\partial F}{\partial s}=\frac{\partial}{\partial s} \Big[ s \Big( \eta_\bk \frac{\partial P}{\partial s}+ \gamma_\bk P\Big)\Big] \label{concl4},
\ee
\begin{align}
\eta_\bk&\doteq192\pi\sum_{\ul{\sigma}}\int d^d{\bk}_2 d^d{\bk}_3 d^d{\bk}_4 \delta^\bk_{234} \delta\left(\tw^\bk_{234}\right) \left|\cH^{-\sigma_2\sigma_3\sigma_4}_{\bk234}\right|^2 \nonumber \\
&\qquad\qquad\qquad\quad \btimes n({\bk}_2) n({\bk}_3) n({\bk}_4) \ge 0, \label{eta}\\
\gamma_\bk&\doteq192\pi\sum_{\ul{\sigma}}\int d^d{\bk}_2 d^d{\bk}_3 d^d{\bk}_4 \delta^\bk_{234} \delta\left(\tw^\bk_{234}\right) \left|\cH^{-\sigma_2\sigma_3\sigma_4}_{\bk234}\right|^2 \nonumber \\
& \btimes \Big[\sigma_2  n({\bk}_3) n({\bk}_4)+\sigma_3  n({\bk}_2) n({\bk}_3) \nonumber 
 +\sigma_4  n({\bk}_2) n({\bk}_3)\Big] \label{gamma}
\end{align}
The conservation equation for $P$ explicitly expresses $F$, the flux of the 1-mode probability in the amplitude space. 
This is a nonlinear Markov evolution equation in the sense of McKean. 
As a matter of fact, the solutions must satisfy a set of self-consistency conditions:
$n(\bk,\tau)=\int ds \;s P(s,\tau;\bk) \label{consistency}$,
where $n(\bk,\tau)$ is the spectrum, that also appears in the formulas for the coefficients (\ref{eta}).
The derivation of the standard kinetic equation from equation (\ref{concl4}) is straightforward. Let us assume that the wave turbulence picture is valid for $s\in(0,s_{nl})$, where the upper bound of the interval can also be $+\infty$ (a fact that will be discussed later). Using (\ref{concl4}), the definition of the wave spectrum $n(\bk)=\int_0^{s_{nl}} s P(s) ds\,$ and integrating by parts, we obtain
\begin{equation}\label{derivkin}
	\frac{\partial n}{\partial \tau} =   \eta_\bk - \gamma_\bk n  - s_{nl} (F(s_{nl}) + \eta_\bk P(s_{nl})).
\end{equation}
The last term is a null term that has to vanish in order for the equation to be satisfied in general, giving a boundary condition in the amplitude space at $s=s_{nl}$. What we are left with is nothing but the kinetic equation. To make it clear for a concrete example of a $4$-wave resonant system where not only $2$ waves $ \rightarrow 2$ waves interactions are present, we derive the kinetic equation for the vibrating plates \cite{Dur_06}. Writing (\ref{derivkin}) in the $2$-dimensional case, we obtain
\begin{align}\label{kineticfull}	
	\frac{\partial n}{\partial\tau}   = 192 \pi &\,\sum_{\ul{\sigma}}  \int d^2 \bk_1 d^2 \bk_2 d^2 \bk_3 \delta^{(2)\bk}_{123} \delta(\tw^{\bk}_{123}) |\cH^{-\sigma_2\sigma_3\sigma_4}_{\bk\bk_1\bk_2\bk_3}|^2 \nonumber \\
	& \times n_\bk n_1 n_2 n_3 \cdot \Big( \frac{1}{n_\bk} + \frac{\sigma_1}{n_1} + \frac{\sigma_2}{n_2} + \frac{\sigma_3}{n_3} \Big),
\end{align}
which is the same equation as in \cite{Dur_06,during2017wave}: the quantity $J_{-\bk\bk_1\bk_2\bk_3}$ in \cite{Dur_06} corresponds to $4i \cH^{-\sigma_2\sigma_3\sigma_4}_{\bk\bk_1\bk_2\bk_3}$  because of the way their coefficients relate to the Hamiltonian coefficients. Therefore, a factor $16$ appears making the two equations identical.
The equation for the pdf can be written also as the following set of stochastic differential equations
\be ds_\bk=(\eta_\bk-\gamma_\bk s_\bk) d\tau + \sqrt{2\eta_\bk s_\bk}dW_\bk, \ee
interpreted in the Ito sense and with self-consistent determination of $n(\bk,\tau)$.
An important solution of (\ref{concl4}) is the  distribution
\be 
Q(s,\tau;\bk)=\frac{1}{n(\bk,\tau)} e^{-s/n(\bk,\tau)}~.
\label{concl5}
\ee
In absence of forcing and dissipation, an H-theorem and the \textit{law of large-numbers} for the empirical spectrum imply that 
the solution relaxes to $Q$, for typical initial wavefields~\cite{Eyi_12,chibbaro20164}. It strictly describes thermodynamic equilibrium only when $n$ is stationary, but our results show (see fig.\ref{fig1}) that $P$ tends to the asymptotic state $Q$ before $n$ has reached its stationary state. This justifies that $Q$ be called distribution of equilibrium despite its formal dependence on time. Furthermore, the results in fig.\ref{fig2} suggest that relaxation to equilibrium also extends to forced and damped systems.

The general stationary solution to eq.(\ref{concl4}) reads~\cite{choi2005anomalous, nazarenko2011wave}
\be 
P(s)=C e^{-s/\nu}-\frac{F_*}{\eta_\bk} \operatorname{Ei}\Big(\frac{s}{\nu}\Big)e^{-s/\nu} \label{gensol}
\ee
where $\operatorname{Ei}(x)$ is the integral exponential function $\operatorname{Ei}(x)=-\int_{-x}^{\infty} \frac{{\rm e}^{-t}}t\,\mathrm dt$. Eq.(\ref{gensol}) is obtained enforcing a constant probability flux in amplitude space: $F(s)=-s\big(\eta_\bk\frac{\partial P}{\partial s}+\gamma_\bk P\big)\equiv F_*$. For the positivity of $P(s)$ for $s\gg \nu$, $F_*$ must be negative, corresponding to a probability flux from the large to the small amplitudes. This must be physically motivated by the existence of strong nonlinear interactions (e.g. breaking of wave crests) which feed probability into the weak, near-Gaussian background. In this picture, this happens at $s=s_{nl}$ and due to the strong nonlinear effects $P(s)$ decays very quickly for $s>s_{nl}$. Thus, the cut-off amplitude $s_{nl}$ and the stationary flux $F_*$ are two aspects of the same phenomenon, connected to each other through the boundary condition that comes out of (\ref{derivkin}) in a natural way:
\be\label{nat}
	P(s_{nl})=-F_*/\eta_\bk.
\ee
This is consistent with the fact that if the weak-turbulence assumption holds over the whole amplitude space, $s\in(0,\infty)$, the normalization of probability implies $F_*=0$, and the equilibrium  exponential distribution is recovered, as expected in absence of strong nonlinear effects that would affect the dynamics. So, clearly the picture with cut-off is meant to describe systems where forcing and damping are present at some wave numbers, which are necessary to sustain the strong nonlinear phenomena. Then, the corrective term in (\ref{gensol}) represents the increased probability in the tail of the distribution due to such nonlinear phenomena ($\operatorname{Ei}(x)\propto \frac{1}{x}$ for $x \gg1$).

Before numerically verifying this scenario, some remarks are in order.
At variance with previous studies\cite{choi2005anomalous, Eyi_12}, we do not need a probability sink to allow the solution,
because we have $F(s)=F_*$ for $s\in (0, s_{nl})$ (similarly as in \cite{nazarenko2011wave}). 
Integrating (\ref{concl4}) from $0$ to $s_{nl}$,
$\frac{\partial}{\partial t}\int_{0}^{s_{nl}} ds P(s)=F(s=s_{nl}) - F(s=0)= 0\label{norm}$,
it is seen that the normalization of the probability in the system is preserved. This appears natural when considering the logarithmic variable $\sigma=\ln(s)$, whose probability density $\Pi(\sigma)$ satisfies
\be \label{simple}
	\partial_t\Pi = \partial_\sigma F,
\ee
with the same $F$ of Eq.~(\ref{concl4}). Imposing $F(s=0)=F_*$, as in the rest of the interval, just means that there is a probability flux from $\sigma_{nl} = \ln(s_{nl})$ toward $\sigma=-\infty$, with probability transferred to infinitesimally small amplitudes.
In the stationary state, using (\ref{nat}) and normalizing the probability yields:
\be\label{normconst}
	C = \frac{1}{\nu} \bigg( 1 + \frac{\Gamma + \ln\frac{s_{nl}}{\nu} - e^{-\frac{s_{nl}}{\nu}}\operatorname{Ei}\big(\frac{s_{nl}}{\nu}\big)}{e^{\frac{s_{nl}}{\nu}}-\operatorname{Ei}\big(\frac{s_{nl}}{\nu}\big)}\bigg)^{-1},
\ee
where $\Gamma \simeq 0.5772$ is the Euler-Mascheroni constant, and $ P(s)=\frac{1}{\nu} e^{-s/\nu} \lb{asympt}$, in the 
$s_{nl} \rightarrow\infty$ limit.
As $s_{nl}$ becomes finite, the complete solution has to be chosen (with $F_*<0$) and this contribution brings a correction to the asymptotic solution.
In conclusion, given the cut-off value $s_{nl}$, which enters as a parameter of the model, and the spectrum $\nu=\eta/\gamma$ 
in the equilibrium limit, the two free constants in (\ref{gensol}) are fixed and a unique general 
solution with cut-off is obtained.

\textit{Numerical results}
In order to validate these analytical predictions, we performed numerical simulations for two
prototype equations of 4-wave turbulence. The first is the Nonlinear Schr\"{o}dinger equation (NLSE) in two dimensions, modeling for instance 
the propagation of electromagnetic fields in optic fibers~\cite{Dyachenko-92}:
\begin{equation}
i \partial_t \Psi =-\frac12 \Delta \Psi +|\Psi|^2 \Psi,
\label{NLSE}
\end{equation}
where $\Delta=\partial_x^2+\partial_y^2$ is the Laplacian operator and $\Psi$ is a field taking complex values.
The second is the F\"{o}ppl Von-Karman equation in two space dimensions for the vibrations of elastic 
plates~\cite{landau}, which in dimensionless form reads:
\begin{eqnarray} 
\frac{\partial^2 \zeta}{\partial t^2} &=& - \frac{1}{4}\Delta^2\zeta +
\{\zeta,\chi\}  ;
\label{foppl0}\\
\Delta^2\chi &=&- \frac{1}{2}\{\zeta,\zeta\}.
\label{foppl1}
\end{eqnarray}
$\chi$ is the Airy stress function imposing the compatibility condition for the displacement field and the Poisson bracket $\{\cdot,\cdot\}$ is defined by
$\{f,g\}\equiv f_{xx}g_{yy}+f_{yy}g_{xx}-2f_{xy}g_{xy}$, so that $\{\zeta,\zeta\}$ is the Gaussian curvature.

The reason for investigating these two models is that they exhibit an important difference in the 4-wave interactions: 
while the NLSE only allows a $2$ waves $\rightarrow 2$ waves collision kernel, because of an additional conservation law, 
the FVK equation allows $1$ wave $\rightarrow 3 $ waves collisions as well. Both equations are solved in a periodic square domain using 
similar numerical schemes involving a pseudo-spectral method (see for instance~\cite{Dur_06} for details on the numerical methods).
We first investigate the evolution of the fields starting with a Guassian distribution (consisting for NLSE of $|\psi({\bf k},0)|^2\propto e^{-k^2/k_0^2}$ with a random phase): the initial pdf of the amplitudes is given by $P(x)=\delta(x-1)$ for each mode, where $x=s/n(0)$ is the normalized amplitude. 
The evolution of the one mode pdf is shown in fig.\ref{fig1} together with the time evolution
of the density spectrum (inset).
We can see that $P(x)$ converges rapidly to the exponential solution given by eq.(\ref{concl5}), in agreement with the theory. Interestingly, the dynamics of the spectrum is different. The spectrum converges towards the 
equilibrium solution given by the Rayleigh-Jeans spectrum\cite{falkovich1992kolmogorov}, but the characteristic time 
is much larger: the pdf has reached equilibrium when the spectrum is still far from it.
That validates the theory and in particular it supports the RPA approximation, which appears 
to be verified from whatever initial conditions after extremely short times.
The same 
dynamics was also observed for the elastic plate (not shown here).
This evidence confirms the results already obtained for a general 3-waves system~\cite{tanaka2013numerical}. 
Then, we study the non-equilibrium wave turbulence energy cascade for the elastic
plate dynamics obtained by injecting energy at large scale through a random noise in Fourier space at small $k$ and
a dissipation dominant at small scale. The balance between these two contributions leads to a stationary regime with a wave turbulence
spectrum following roughly $ |\zeta_k|^2 \sim k^{-4}$ at low forcing (up to a logarithmic correction~\cite{Dur_06}) that 
corresponds to a constant flux of energy from the large to the small scales. It is thus tempting
to compare the pdf of the Fourier modes of this dynamics with that of the Hamiltonian 
dynamics studied above, for which the theory has been derived. Indeed, no theoretical predictions can be 
easily made in such configuration, because the forcing-dissipation terms break the Hamiltonian structure. 
Moreover, while a distribution 
close to the one of the equilibrium situation could be expected at low forcing, intermittency at high 
forcing is supposed to heavily influence the pdf of the Fourier mode, similarly to what has been observed for the
high moments of the structure function in real space~\cite{chibbaro2015elastic}. Surprisingly,
fig.~\ref{fig2} shows that the pdf's are very close to the Rayleigh distribution predicted for the Hamitonian 
dynamics, in the absence of flux ($F_*=0$) even at high forcing where the spectrum exhibits a $k^{-6}$ slope at small $k$. However, a closer analysis shows a slight deviation from this
distribution for modes at small scales, just before the dissipative range, where the pdf is
better fitted by the generalized distribution (\ref{gensol}) with $F_*\neq0$. Similar results have also been observed for the NLSE with no noticeable non-zero $F_*$. The weak value of $F_*$ obtained for our systems suggests that while clear signature of intermittency is detected in physical space via structure functions\cite{chibbaro2015elastic},
it is difficult to find anomalous scaling looking at the 1-mode spectral pdf.
On one hand, the effect is expected to be small for those systems where the spectrum of wave turbulence is only a small logarithmic correction to the equilibrium spectrum, so that the dominant signal in the fluctuations of the spectrum is due to the statistical equilibrium contribution. This is certainly the case for NLSE. On the other hand, fig.\ref{fig2} suggests a non-trivial interplay between large and small scales, since in vibrating plates the spectrum is definitely far from equipartition at large-scales, but signature of intermittency is found at very small scales, even in physical space\cite{chibbaro2015elastic}. This issue deserves future investigation.

\textit{Acknowledgements}
The authors gratefully acknowledge the referee's
insightful remarks, that also allowed them to correct one error.

\bibliography{references}

\end{document}